\documentclass{ws-ijmpd}
\usepackage[super,compress]{cite}
\begin{document}
\baselineskip=12pt
\def\black{\textcolor{black}}
\def\red{\textcolor{black}}
\def\blue{\textcolor{blue}}
\def\green{\textcolor{black}}
\def\be{\begin{equation}}
\def\ee{\end{equation}}
\def\bea{\begin{eqnarray}}
\def\eea{\end{eqnarray}}
\def\orc{\Omega_{r_c}}
\def\om{\Omega_{\text{m}}}
\def\E{{\rm e}}
\def\bearst{\begin{eqnarray*}}
\def\eearst{\end{eqnarray*}}
\def\peleven{\parbox{11cm}}
\def\peffec{\peight{\bearst\eearst}\hfill\peleven}
\def\pspace{\peight{\bearst\eearst}\hfill}
\def\ptwelve{\parbox{12cm}}
\def\peight{\parbox{8mm}}
\markboth{Habibi, Baghram, Tavasoli }
{Peculiar velocity measurement in a clumpy universe }

%
\catchline{}{}{}{}{}
%

\title{Peculiar velocity measurement in a clumpy universe  }

\author{FARHANG HABIBI}

\address{LAL, IN2P3-CNRS, \\
 B.P. 34, 91898 Orsay Cedex,
France\\
habibi@lal.in2p3.fr}

\author{SHANT BAGHRAM}

\address{Department of Physics, Sharif University of
Technology,\\ P.~O.~Box 11155-9161, Tehran, Iran\\
baghram-AT-sharif.edu}

\author{SAEED TAVASOLI}

\address{Department of Physics, Kharazmi University,\\
P.~O.~Box 31979-37551, Tehran, Iran\\
tavasoli@ipm.ir}

\maketitle
%

\begin{abstract}
{\bf{Aims:}} In this work we address the issue of peculiar velocity measurement in a perturbed Friedmann universe using the deviations from measured luminosity distances of standard candles from background FRW universe. We want to show and quantify the statement that in intermediate redshifts ($0.5< z < 2$),  deviations from the background FRW model are not uniquely governed by  peculiar velocities. Luminosity distances are modified by
gravitational lensing. We also want to indicate the importance of relativistic calculations for peculiar velocity measurement at all redshifts.
 \\  {\bf{Methods:}} For this task we discuss the relativistic correction on luminosity distance and redshift measurement and show the contribution of each of the corrections as lensing term, peculiar velocity of the source and Sachs-Wolfe effect. Then we use the SNe Ia sample of Union 2, to investigate the relativistic effects we consider. 
 \\{\bf{Results:}} we show that, using the conventional peculiar velocity method, that ignores the lensing effect, will result in an overestimate of the measured peculiar velocities at intermediate redshifts. Here we quantify this effect. We show that
at low redshifts the lensing effect is negligible compare to the effect of peculiar velocity.
From the observational point of view, we show that the uncertainties on luminosity of the present SNe Ia data prevent us
from precise measuring the peculiar velocities even at low redshifts ($z<0.2$).
\end{abstract}




\section{Inroduction}
The propagation of the light from  standard candles, such as  supernovae type Ia (SNe Ia),
from the source to the observer is  of  greatest interest for  cosmology.
On one hand, standard candles are used to measure the rate of the expansion of the universe
\cite{Riess:1998cb,Perlmutter:1998np}. On the the other hand, the amount of light bundle
distortion of the source is used to measure the gravitational effect of the matter
distribution along the line of sight, which helps us to learn about the distribution and evolution of the
matter in different cosmic times and scales\cite{Bartelmann:1999yn,Bonvin:2005ps}. Another phenomenon which has an important effect
on the measurement of the magnitude  of the standard candle and its interpretation is the peculiar velocity. This is because the peculiar velocity changes the redshift of the source \cite{Davis:2010jq}. The host galaxies of standard candles has an additional velocity beside
the Hubble expansion velocity. Although the existence of peculiar velocity makes ambiguity
in the measurement of the cosmological redshift, however it is used as a  promising tool
to probe the matter distribution, this is because the peculiar velocity of matter is sourced by the change in the matter perturbation due to continuity equation. Accordingly the peculiar velocity of luminous matter (i.e. galaxies) can be used as a tracer of the
dark matter distribution in the Universe. As future large-scale structure surveys will
open up a great opportunity to map the Universe on larger scales and at deeper redshifts,
peculiar velocity measurements, lensing maps of the Universe and distance
measurements of standard candles  become progressively more important\cite{Koda:2013eya}. These observations can
be used as a probe to check the standard model of cosmology
or to see any deviations from $\Lambda$CDM \cite{Baghram:2009fr, Baghram:2010mc, Baghram:2014qja , Khosravi:2015boa}. Accordingly, the
accurate measurement of these quantities, their hidden relations and their model
dependencies have great importance. Historically, the peculiar velocities are obtained
by the relation {$v_p=cz - Hr$, where $v_p$ is the peculiar velocity, $z$ is the observed
redshift, $H$ is the value of Hubble parameter and $r$ is the physical distance
to the source respectively and $\bar{v}=Hr$ is the velocity of the Hubble flow
\cite{Kaiser:1989kb,Dekel:1994sx,Strauss:1995fz,Masters:2006tb,Sarkar:2006gh,Abate:2008zy,Lavaux:2007zw,Nusser:2011tu}.
Using the peculiar velocity relation one can measure the distances to the
sources and having the observed redshifts, the peculiar velocity can be deduced.
However, distance measurement is a non-trivial exercise and there are many different methods to do this.
The Tully-Fisher relation for spirals\cite{Tully:1977fu}, surface brightness fluctuations for nearby galaxies\cite{Tonry:2000aa},
the Faber-Jackson relation for elliptical galaxies\cite{Faber:1976sn},
the tip of the Red Giant Branch, Cepheids, and SNe Ia are the most developed methods to measure the distances\cite{Freedman:2000cf}.
In this work, first we reexamine the luminosity distance measurements in a clumpy universe by accounting for the lensing \cite{Holz:2004xx,Hilbert:2008kb,BenDayan:2013gc,Bolejko:2012ue},
Sachs-Wolfe and Integrated Sachs-Wolfe effects \cite{Bacon:2014uja}. Then we demonstrate the relativistic
approach to measuring peculiar velocities similar to the one  presented in Davis and Scrimgeour \cite{Davis:2014jwa}, and  finally we take into account both relativistic corrections and lensing corrections to luminosity distance computation in $\Lambda$CDM framework.
In this work, we also study the physics of the peculiar velocity measurement, where we show that the distance measurement methods which assume that the observed  distance to a host galaxy is equal to its proper distance are all biased. This bias comes from the fact that deviation from unperturbed FRW Universe
is not considered in distance measurement. In other words, the over/under dense regions along the line of sight affect the
luminosity of the standard candle.
Accordingly we assert that the distance-independent measurements, like peculiar velocity measurements by redshift-space distortions, linear theory or kinetic Sunyaev-Zeldovich effect  are more unbiased observation to find the peculiar velocity.
Although with the nowadays data, we currently have larger errors using these methods\cite{Ma:2014lua},
the results of this study will be important to be applied in future Large Scale Structure (LSS) surveys, which probe the Universe in higher redshifts,
where the unbiased measurement of the distances become important. It is very essential to note that the full relativistic treatment of the peculiar density measurements can be done by  the use of spacetimes equipped with two families of observers,(one which is fixed with CMB reference frame and the other which follow the matter (like a galaxy) flow) with their associated 4-velocities forming a "tilt angle".\cite{Ellis:2002tq}
In this direction there are studies that shows how peculiar velocities could change the way observers interpret their cosmological data \cite{Tsagas:2009nh, Tsagas:2011wq, Tsagas:2015mua}. However in this work we do not use the tilted coordinates, which can be  an idea for extending this study.
The structure of this work is as follow: In the first subsection of Sec.(\ref{Sec2}) we study the Luminosity distance in clumpy Universe without peculiar velocity. In the second subsection we study the case of peculiar velocity as a redshift redefinition parameter and in the third subsection we take into account the physics of both peculiar velocity and gravitational lensing. In Sec.(\ref{Sec3}) we have the conclusion and future prospects.



\section{Luminosity distance}
\label{Sec2}
In a homogenous-isotropic universe with FRW metric:
\be
ds^2=-dt^2+a^2(t)d\sigma^2
\ee
where $d\sigma^2=d\chi^2 +f^2_K(\chi)[d\theta^2+\sin^2\theta d\phi^2]$ is the spatial line element and $f_{K}$ is:
\be
f_K(\chi) = \begin{cases} \sin\chi &\mbox{if } K=+1 \\
\chi & \mbox{if } K=0 \\
\sinh\chi & \mbox{if } K=-1 \\
 \end{cases}
\ee
where $\chi$ is the comoving distance and $K$ is the space curvature.
We can measure the cosmological parameters by  measuring the luminosity
distance of the standard candles such as SNe Ia.
The luminosity distance $d_L$ is defined as the ratio of the absolute luminosity $L_s$ of the source to the observed flux  ${\cal{F}}$, by:
\be
d_L^2=\frac{L_s}{4\pi {\cal{F}}}.
\ee
The luminosity distance in the flat-FRW model is related to comoving distance by:
\be  \label{eq:dL}
\bar{d_L}(\bar{z})=(1+\bar{z})\chi(\bar{z})=(1+\bar{z})\int_0^{\bar{z}}\frac{c dz'}{H(z')},
\ee
where $\bar{d_L}(\bar{z})$ is the background luminosity distance (luminosity distance in the unperturbed universe described
by FRW metric) at background (or cosmological) redshift $\bar{z}$. Note that hereafter $\bar{}$ indicates the background quantity.
The background luminosity distance is related to the distance modulus by:
\be
\bar{\mu}=m-M= 5\log (\bar{d_L}(\bar{z})) +25,
\ee
where $m$ and $M$ are the apparent and absolute magnitudes of a standard candle (such as SNe Ia) and in this case
$d_L$ is measured in Mpc units and $\mu$ can be interpreted as a distance indicator
(amount of magnitude change), interchangeably.

In the first subsection, we study the distance  measurement  case, in a clumpy universe with change in the background redshift. In the second subsection, we study the case of unperturbed Universe with the inclusion of the effects that change the observed background redshift like peculiar velocity. This will be done in relativistic manner and finally in the third subsection we study both the effect of peculiar velocity  and convergence on the luminosity distance measurement.


\subsection{Luminosity distance in a clumpy universe with zero peculiar velocities}
\label{subsec:A}
The Universe is not exactly homogenous and isotropic. The cosmic structures include over-dense (galaxies, groups and clusters of galaxies)
and under-dense (voids) regions which makes the Universe to differ from being a FRW space-time. Hence, we should consider a perturbed metric for the clumpy Universe in order
to investigate the propagation of light through voids and over-dense regions. We use the perturbed FRW-metric in Newtonian gauge:
\be
ds^2=-(1+2\Psi(\vec{x},t))c^2 dt^2 + a^2(t) (1-2\Phi(\vec{x},t)) d\sigma^2,
\ee
where $\Psi(\vec{x},t)$ and $\Phi (\vec{x},t)$ are the perturbed scalars.
In the case that we have the general relativity as the classical theory of gravity and the assumption that the cosmic fluid does not have  an anisotropic terms in its energy-momentum tensor, we have $\Psi=\Phi$.

In order to study the effect of inhomogeneities on the propagation of the light bundle we have to solve the Sachs equation \cite{Bartelmann:1999yn}:

\be
\frac{d^2}{d\chi^2}d_L=-\frac{1}{2}R_{\mu\nu}k^{\mu}k^{\nu}d_L,
\ee
where we replaced the angular diameter distance with luminosity distance as in metric theories
we have $d_{L}=(1+z)^2d_{A}$. $R_{\mu\nu}$ is the Ricci tensor and $k^\mu$ is the four momentum of photon which is defined by using the background comoving distance
as the affine parameter $k^\mu=dx^\mu/d\chi$.
Using the Sachs equation the luminosity distance in perturbed universe $d_L(\bar{z})$ becomes:
\be
\label{debu}
d_L(\bar{z})=\bar{d}_L(\bar{z})\left[1+\frac{\delta d_L}{\bar{d}_L(\bar{z})}\right],
\ee
where ${\delta d_L}/{\bar{d}_L(\bar{z})}$  is the change of the luminosity distance with respect to the FRW prediction, due to effect of structures along the line of sight.
The ${\delta d_L}/{\bar{d}_L(\bar{z})}$ is as below \cite{Bacon:2014uja}:

\be \label{eq:deltadl}
\frac{\delta d_L}{\bar{d}_L}|_{\chi_s}=\hat{n}.\vec{v}_0/c-\Phi_s + \frac{1}{\chi_s}\int_0^{\chi_s}d\chi_s(2\Phi - (\chi-\chi_s)\chi\nabla^2 _{\perp}\Phi),
\ee
where $\vec{v}_0$ is the peculiar velocity of the observer, $\hat{n}$ is the unit vector from the observer toward the source, $\Phi_s$ is the gravitational potential of the host and $\chi_s$ is the comoving distance to source respectively. The 2 dimensional Laplacian is defied as ($\nabla^2_{\perp}=\nabla^2-(\hat{n}.\vec{\nabla})^2+2\hat{n}.\vec{\nabla}/\chi$). It is worth to mention that there is no term related to the peculiar velocity of source in luminosity distance change obtained Eq.(\ref{eq:deltadl}). The other important thing to indicate is that a crucial term in the luminosity distance change is the lensing convergence $\kappa_g$ term, which is defined from the second term of the integral in Eq.(\ref{eq:deltadl}) as:
\begin{eqnarray}  \label{kgformula}
\kappa_g &=&\int_0^{\chi_s}d\chi (\chi_s - \chi)\frac{\chi}{\chi_s}\nabla_{\perp}^2\Phi   \\ \nonumber & \simeq & \frac{3}{2}H_0^2\Omega_m\int_0^{\chi_s}d\chi(\chi_s-\chi)\frac{\chi}{\chi_s}(1+z_s)\delta_m(\chi),
\end{eqnarray}
where the second equation is an approximation because we exchange the 2D Laplacian with that in 3D.
We use the Poisson equation to replace the $\Phi$ with density contrast $\delta_m$, which depends on the comoving distance.
$\Omega_m$ is the matter density parameter in the present time.








In the case that we can neglect the potential terms with respect to the convergence term, the luminosity distance of the standard candles will be modified as below:
\be
\label{eq:kg}
d_L(\bar{z})=\bar{d}_L(\bar{z})[1-\kappa_g(\bar{z})],
\ee
where $d_{L}(\bar{z})$ is the observed luminosity distance at comoving redshift,  $\bar{z}$, produced by the Hubble expansion.
The $\kappa_g$ is the gravitational lensing correction obtained from the line of sight integration of the matter density contrast as in Eq.(\ref{kgformula})\cite{Bartelmann:1999yn}.
We should note that gravitational lensing only change the source apparent magnitude and do not affect
the redshift of the source. Therefore, in the absence of peculiar velocities,
the observed redshift is the same as the comoving redshift: $z=\bar{z}$.

An important example of a host galaxy with no peculiar velocity is the one that resides at the center of
a cosmic void or center of a galaxy cluster (a Brightest Central Galaxy (BCG)).



In addition, in relation (\ref{eq:deltadl}), there are two  gravitational potential terms other
than $\kappa_g$ that can affect the observed luminosity. In the next section, we will include
the effect of gravitational potential change on the source luminosity and will show
that at low and intermediate redshifts this effect is negligible.


\subsection{Luminosity distance with peculiar velocity as an example of redshift redefinition}
\label{subsec:B}

\begin{figure}
\centerline{\psfig{file=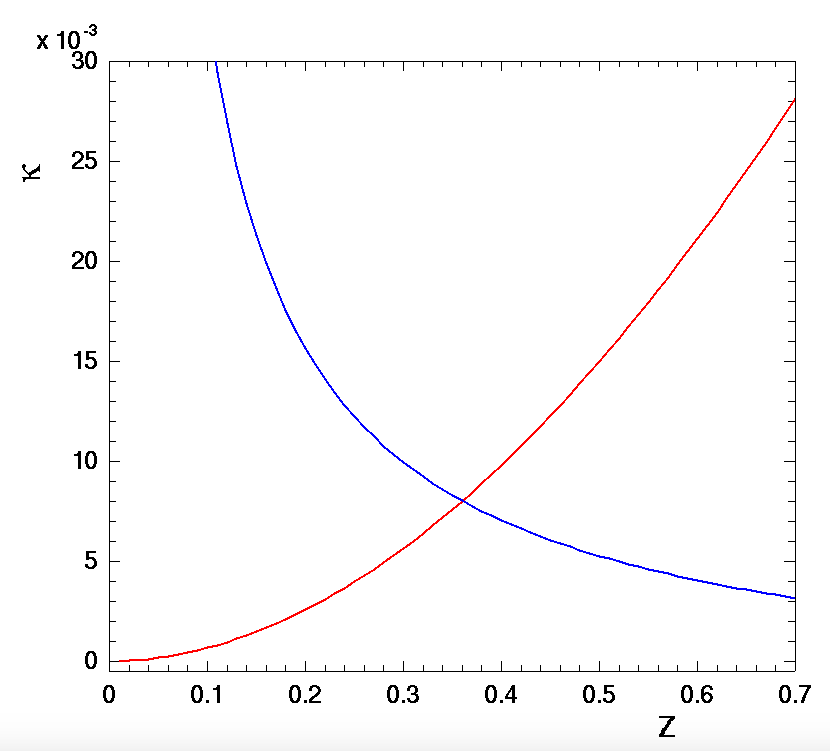,width=7.7cm}}
\vspace*{8pt}
\caption{Red curve shows the change of the theoretical variance of $\kappa_g$ versus redshift. Blue curve represents
the variation of $\kappa_v$ for $v_p$=1000 km/s as a function of redshift. Gravitational lensing stars dominating the peculiar velocity
after $z \sim$ 0.35. The curves are plotted for a flat $\Lambda$CDM universe with $\Omega_m$=0.27 and $H_0$=70 km/s/Mpc.}
\label{kg04}
\end{figure}

At  low redshifts ($z<0.1$), the line-of-sight  effects do not significantly affect the light bundle of the standard candles.
 { In order to confirm this assertion, { we consider the fact that}
 the variance of the $\kappa_g$ is related to the matter power spectrum via Eq.(\ref{kgformula}).
 We plot this theoretical variance of $\kappa_g$ versus redshift in Fig. \ref{kg04}.
}
 {According to the figure, at low redshifts $\kappa_g \sim 10^{-3}$. By using relation (\ref{eq:kg}),
this creates a typical distance modulus fluctuation of $\sim 10^{-3}$ magnitude.}
 As we will see in this section, this
fluctuation is negligible in comparison to the fluctuation produces by peculiar velocity which is at order of 0.1 magnitude.
The peculiar velocities of the sources, produced by the local density contrast (as in  the case of a SNe Ia host galaxy which resides
in in-falling regions in a galaxy group ), can have a dominant effect on the luminosity distance computed by background $\Lambda$CDM model
through changing the redshift from $\bar{z}$ to $z$. This redshift distortion can be \emph{corrected} to reproduce the
background model by Taylor expansion of Eq.(\ref{eq:dL}) around the observed redshift $z$:
\be
\bar{d}_L(\bar{z})=\bar{d}_L(z)+\frac{\partial \bar{d}_L}{\partial \bar{z}}|_{z}(\bar{z}-z).
\label{dlbarv}
\ee
The non-zero $\bar{z}-z$ is produced by the peculiar velocity of the source.
However in order to take all the terms that contribute to the redshift redefinition, we have to solve the volume expansion
equation \cite{Bartelmann:1999yn}:
\be\label{eq:expansion}
\frac{dz}{d\chi}=[-\frac{\Theta}{3}+\sigma_{\mu\nu}n^\mu n^\nu](1+z)^2
\ee
where $\Theta$ is the expansion parameter, $\sigma_{\mu\nu}$ is anisotropic stress and $n^{\mu}$ is normal 4-vectors.
Solving the Eq.(\ref{eq:expansion})we will get:
\be{\label{eq:deltazz}}
\frac{z-\bar{z}}{1+\bar{z}}=z_p - z_{SW}/2 - 2\int_0^{\chi_s}d\chi \Phi',
\ee
where $z_p$ is the peculiar redshift due to the peculiar velocity of the source and $z_{SW}$ = $2 (\Phi_s-\Phi_0)$
is the redshift induced by Sachs-Wolf effect. The gravitational potential can be estimated as
$\Phi \sim (v_s/c)^2$ where $v$ is the dispersion velocity of the source host galaxy. At intermediate and low redshifts
 $z_{SW} \ll z_p$.

\begin{figure}
\centerline{\psfig{file=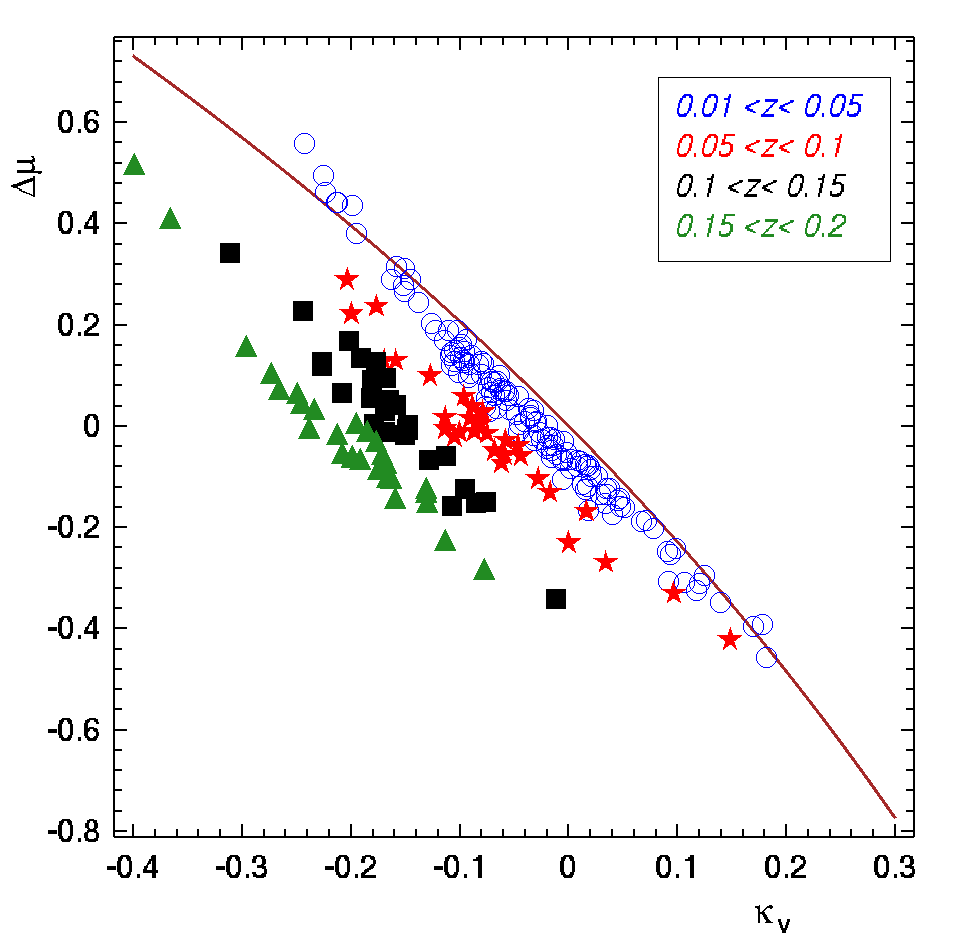,width=7.7cm}}
\caption{The solid curve represents relation (\ref{muzkv}).
The symbols show peculiar redshifts of SNe computed non-relativistically through relation (\ref{vpapprox2})
and substituted in $\kappa_v$ relation. The deviation from the relativistic curve increases by redshift.}
\label{delmukv}
\end{figure}

Neglecting the gravitational potential contribution,
we focus our attention to the peculiar velocity as the main source of
the redshift deviation from the background redshift.
Using the relativistic peculiar redshift, we have:
\be
1+z_p=\sqrt{\frac{1+(v_p/c)}{1-(v_p/c)}},
\label{zp}
\ee
where $c$ is the light speed.
{$v_p$ is the peculiar velocity along the line of sight which is related to the source proper motion
through relation $v_p = \hat{n}.\vec{v}_s$, where $\hat{n}$ is the unit vector pointing from the observer to the source.}
One should note that the  equation above is valid for measuring any relativistic velocity.
In order to formulate the $v_p$, we start with the definition of physical distance $r=a\chi$, where $a=1/(1+z)$ is the scale factor.
We neglectthe gravitational potential corrections on proper distance measurement.
Consequently, the observed velocity in non-relativistic limit is written as:
\be
\label{vpapprox1}
v=\bar{v}+v_p
\ee
where $\bar{v}$ is the cosmological Hubble velocity ($\bar{v}=H r$) .
In order to relate the peculiar velocity $v_p$ to the observed redshift $z$ and the unperturbed cosmological redshift $\bar{z}$,
we use the relativistic relation \cite{Davis:2014jwa}
\be \label{eq:vz}
v=c\frac{(1+z)^2-1}{(1+z)^2+1},
\ee
which in the non-relativistic limit  reduces to $v\simeq cz$ and gives the  well known but approximate relation:
\be
\label{vpapprox2}
v_p\simeq cz - H d_L(z).
\ee
{where we have approximated $r(z)$ by the luminosity distance $d_L(z)$.}
In \cite{Davis:2014jwa}, they show that the relation $v\simeq cz$
introduces accountable errors even at low redshift.
The correct relation between the redshifts is as below:
\be
\label{redshift}
(1+z)=(1+\bar{z})(1+z_p).
\ee
The relation $z \simeq\bar{z}+z_p$ works only at low redshifts and small peculiar velocities.
The equation (\ref{redshift}) can be derived by considering how two different
inertial observers, located at any co-moving distance from each other on the Hubble flow, measure the
wavelength emitted by a source with a peculiar motion.
The important task now is to find
the peculiar velocity (or equivalently the peculiar redshift $z_p$).
By computing the derivative of equation (\ref{eq:dL}) and substituting $\bar{z}-z$ from equation (\ref{eq:deltazz})
(where we neglect the potential related terms), the equation (\ref{dlbarv}) becomes as:
\be
\label{dlbarv2}
\bar{d}_{L}(\bar{z})=\bar{d}_L(z)\left[1 - \kappa_v \right],
\ee
where by considering the GR corrections described by \cite{Bonvin:2005ps}:
\be
\label{kappav}
\kappa_v =  \left( \frac{c(1+z)}{\chi (z) H(z)} -1 \right)\frac{z_p}{1+z_p}.
\ee

We rewrite relation (\ref{dlbarv2}) in distance modulus form:
\be
\label{mukv}
\bar{\mu}(\bar{z})=\bar{\mu}(z) + 5\log\left[1 - \kappa_v \right].
\ee
One observes the distance modulus, $\mu(z)$, of a standard candle at observed redshift $z$ and
assumes that the measured luminosity belongs to the non-distorted redshift $\bar{z}$ and put
$\mu(z)=\bar{\mu}(\bar{z})$ (redshift distortion does not affect the background luminosity of the source).
We hence re-write the relation (\ref{mukv}) as:
\be
\label{muzkv}
\Delta\mu(z) = 5\log\left[1 - \kappa_v \right],
\ee
where $\Delta\mu(z)= \mu(z)-\bar{\mu}(z)$ is the deviation of the observed luminosity from the
background model at the observed redshift $z$.

It is interesting to note that for a given cosmology and a given peculiar velocity, $\kappa_v$ changes sign at redshift $z_* \sim$ 1.5
At redshifts $z<z_*$, if the source is moving toward us it is observed to be fainter than what model predicts. This will be inverse for the same object
at redshifts $z>z_*$.
 In general, depending on the source redshift and the direction of the source along the line of sight, the source will look brighter or fainter than what the model predicts. Since this effect acts like (de)lensing, it is called Doppler lensing\cite{Bacon:2014uja}.
{By observing the distance modulus $\mu(z)$ and computing $\Delta \mu = \mu - \bar{\mu}$,
the parameter $\kappa_v$ and the peculiar velocity are computed by using equation (\ref{muzkv}) and (\ref{kappav}).}
One can show that relation (\ref{vpapprox2}) is the approximate version of relation (\ref{dlbarv2}) at low redshifts.
In Fig. \ref{delmukv}, the solid curve demonstrates
relation (\ref{muzkv}). The distance modulus of the data points are given by Union 2 SNe Ia catalog \cite{Amanullah:2010}
for $z<0.2$.
$\bar{\mu}$ is computed for $\Omega_m$=0.27, $\Omega_\Lambda$=0.73 and $H_0$=70 km/s/Mpc.
The coloured symboles represent peculiar velocities computed from relation (\ref{vpapprox2}).
We have substituted the non-relativistic $v_p$ in relation (\ref{kappav}) and plotted $\Delta\mu$ vs. $\kappa_v$.
The non-relativistic approximation overestimates the absolute value of the peculiar velocities.
The overestimation increases by redshift.

We can compare the variation of $\kappa_g$ and $\kappa_v$ as function of redshift in Fig. \ref{kg04}.
It shows that at low redshift the lensing effect is subdominant and the luminosity deviation from the background model
is caused mainly by the peculiar velocities.


\begin{figure}
\centerline{\psfig{file=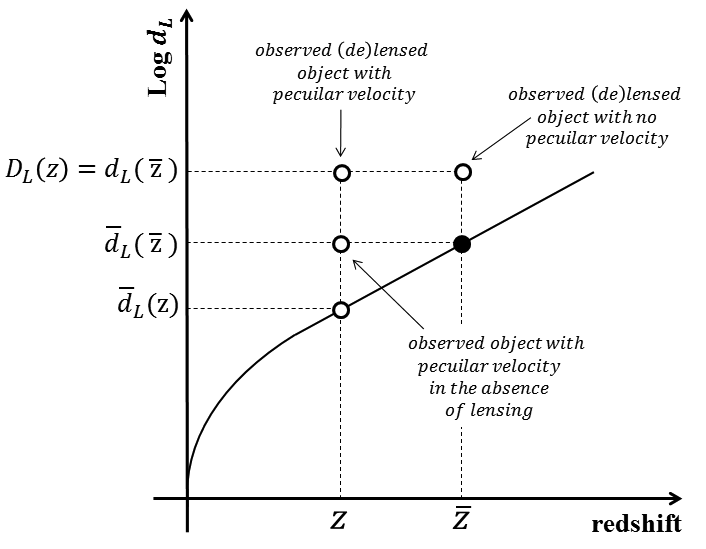,width=7.7cm}}
\caption{Schematic view of the effect of peculiar velocities and gravitational lensing on the background cosmology
curve (solid curve). A standard candle, located originally at redshift $\bar{z}$ with unperturbed luminosity distance $\bar{d}_L(\bar{z})$,
will be measured at a different redshift ($z$), due to peculiar velocity, and different luminosity distance
($D_L$) due to (de)lensing effect. At redshift $z$, we expect a standard candle to have (unperturbed) luminosity distance
$d_L(z)$.}
\label{dlschem}
\end{figure}

%


To include the effect of changes in gravitational potentials on the luminosity and redshift measurement,
we substitute equation (\ref{dlbarv}) in equation (\ref{debu}).
By taking into account the relations (\ref{eq:deltadl}) and (\ref{eq:deltazz}), we obtain \cite{Bacon:2014uja}:
\be
\label{dlztot}
D_L(z)=\bar{d}_L(z)\big[ 1-\kappa_v - \kappa_ g - \kappa _{SW} - \kappa _{ISW}],
\ee
where $D_L(z)$ is the observed luminosity distance at observed redshift $z$, and
 \be
\label{ksw}
\kappa_{SW}(z)=2\Phi_s(z) + \frac{c(1+z)}{2H(z)\chi(z)} z_{SW},
\ee
\be
\label{kisw}
\kappa_{ISW}({z})=\frac{2}{\chi_s({z})}\int_0^{\chi_s({z})}d\chi \Phi + \frac{2}{c} (1 +\frac{c(1+z)}{H(z)\chi(z)})\int_0^{\chi_s(z)}d\chi \Phi'.
\ee
The last equation shows the contribution of the Integrated Sachs-Wolf effect on the luminosity fluctuation.
This term contributes more for sources located at high redshifts. At low and intermediate redshifts ($z<2$),
we neglect $\kappa_{SW}$ and $\kappa_{ISW}$.

\subsection{Luminosity distance with peculiar velocity in a clumpy universe}
Consider an unperturbed FRW Universe where the luminosity
distances of the celestial objects are described by equation (\ref{eq:dL}).
For the sources located at intermediate redshift,
the existence of voids and clumps can perturb  equation (\ref{eq:dL}) in two ways:\\
1. The lensing effect of line-of-sight  structures keeps the comoving redshift unchanged and changes
the source background luminosity, through relation (\ref{eq:kg}). \\
2. The peculiar velocity, induced locally by nearby structures of
the host galaxy, keeps the (de)lensed luminosity unchanged and perturbs the comoving redshift from
$\bar{z}$ to $z$ (Doppler lensing) through relation (\ref{dlbarv2}).
These effects are schematically shown in Fig. \ref{dlschem}. Consider an object located
at comoving redshift $\bar{z}$ with background luminosity distance $\bar{d}_L(\bar{z})$.
{{We observe the same object at redshift $z$ and luminosity distance $D_L(z)$.
From Fig. \ref{dlschem}, we have $D_L(z)=d_L(\bar{z})$ (lensing is the only cause of (de)magnifying the
background luminosity).}} By neglecting the SW and ISW effects, we will have:
\be
D_L(z) = \bar{d}_L(z)\left[1-\kappa_v-\kappa_g  \right],
\label{DLdL}
\ee

We now have all the elements to compute the peculiar velocity at the observed redshift $z$:
\be
v_p/c = (1-\kappa_g - \frac{D_L(z)}{\bar{d}_L(z)})(\frac{c(1+z)}{\chi(z)H(z)}-1)^{-1}
\label{unbvp}
\ee
For $\kappa_g=0$, we arrive at equation (\ref{muzkv}) where $\mu(z)=5\log{D_L(z)}+25$.
Measuring the peculiar redshift from the observed $D_L(z)$ (or $\mu(z)$) by the method described in the previous
section ( Eq.(\ref{muzkv})) will obviously overestimate the peculiar velocity.} This is because
by assuming $\mu(z)=\bar{\mu}(\bar{z})$,
we are attributing all of the deviation from the Hubble diagram to redshift-space distortion, neglecting
the gravitational (de)lensing.

A numerical example of velocity overestimation can be given as follow. It is possible to have a
convergence factor $\kappa_g \sim$ 0.1
at redshift $\bar{z}\sim$ 1. This convergence is equivalent to  distance modulus excess  $|\Delta \mu| \sim$ 0.2.
If one attributes this luminosity excess to peculiar velocity by using Eq. (\ref{muzkv}), we end up with
$v_p \sim$ 10,000 km/s which is unrealistic.

From observational point of view, we can not yet easily separate the contribution of the peculiar velocity and gravitational
(de)lensing. A profound map for $\kappa_g$ is needed to estimate the (de)lensing effect at intermediate redshifts.
This map is not yet available. Moreover, the presence of the systematic errors and uncertainties on the luminosity measurement of
the standard candles can mimic the observed luminosity residuals of the Hubble diagram.

To compute the lensing map, the density contrast field, $\delta_m$, in equation (\ref{kgformula})
should be integrated along the line of sight up to the intermediate redshifts. The method is explained
in \cite{Baghram:2014qja} to compute $\kappa_g$ map with  $z_{max}=$0.04 for SDSS DR10.
Assuming a unity bias between the CDM and luminous matter, $\delta_m$ can be estimated by measuring
the luminosity contrast of galaxies, residing in over and under dense regions. Regarding the void galaxies,
 Tavasoli et al \cite{Tavasoli:2015} explained how to derive the void catalogues
and compute the density contrast by applying a 3D grid on a volume-limited galaxy sample.
One of the parameters that affects the precision on $\kappa_g$ measurement
is the spatial resolution of the derived density contrast field.

According to Union 2 compilation paper \cite{Amanullah:2010},
one of the main sources of the intrinsic dispersion in distance modulus measurement
is the 0.15 magnitudes error assigned to SNe Ia of the whole
sample to decrease the weight of the poorer-measured (sub)samples. Regarding other systematics,
they have shown that the propagation of the calibration uncertainties of the photometric passbands on the
measured distance moduli should be taken into account for each SNe Ia sample separately. This can add
uncertainties up to 0.04 magnitudes depending on the sample and the passband. Considering the
photometric and spectroscopic diversity of the SNe Ia for wavelengths shorter than rest-frame B-band,
they included 0.03 magnitudes correlated uncertainty for all SNe Ia with photometric
band bluer than rest-frame 3500\AA. This affects more the SN Ia at higher redshifts.
Other sources of uncertainties such as Malmquist bias,
Galactic and intergalactic extinction are included as well but they have smaller contributions.
Regarding all these facts, we analyse the luminosity residuals of the Hubble diagram as follow:

\begin{figure}
\centerline{\psfig{file=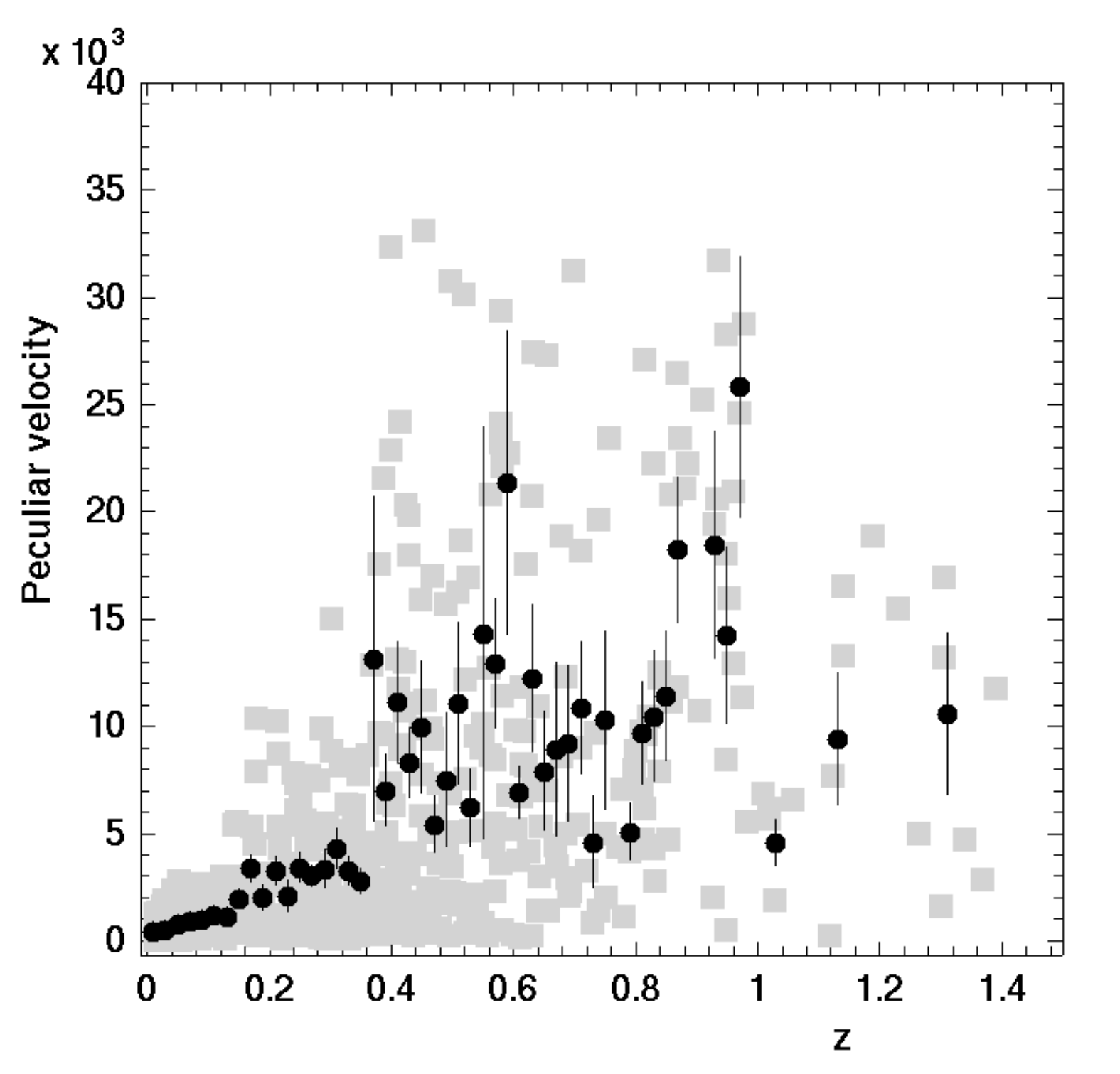,width=7.7cm}}
\caption{The absolute values of the peculiar velocities (km/s), computed from relation (\ref{muzkv}) up to
$z$=1.4 for SNe Ia of Union2 catalog. The data (grey boxes) are averaged with bin width of 0.01 redshift (black circles).
The error bars show one sigma scatter per bin. No correction due to lensing effect is applied.}
\label{2plot}
\end{figure}

\begin{figure}
\centerline{\psfig{file=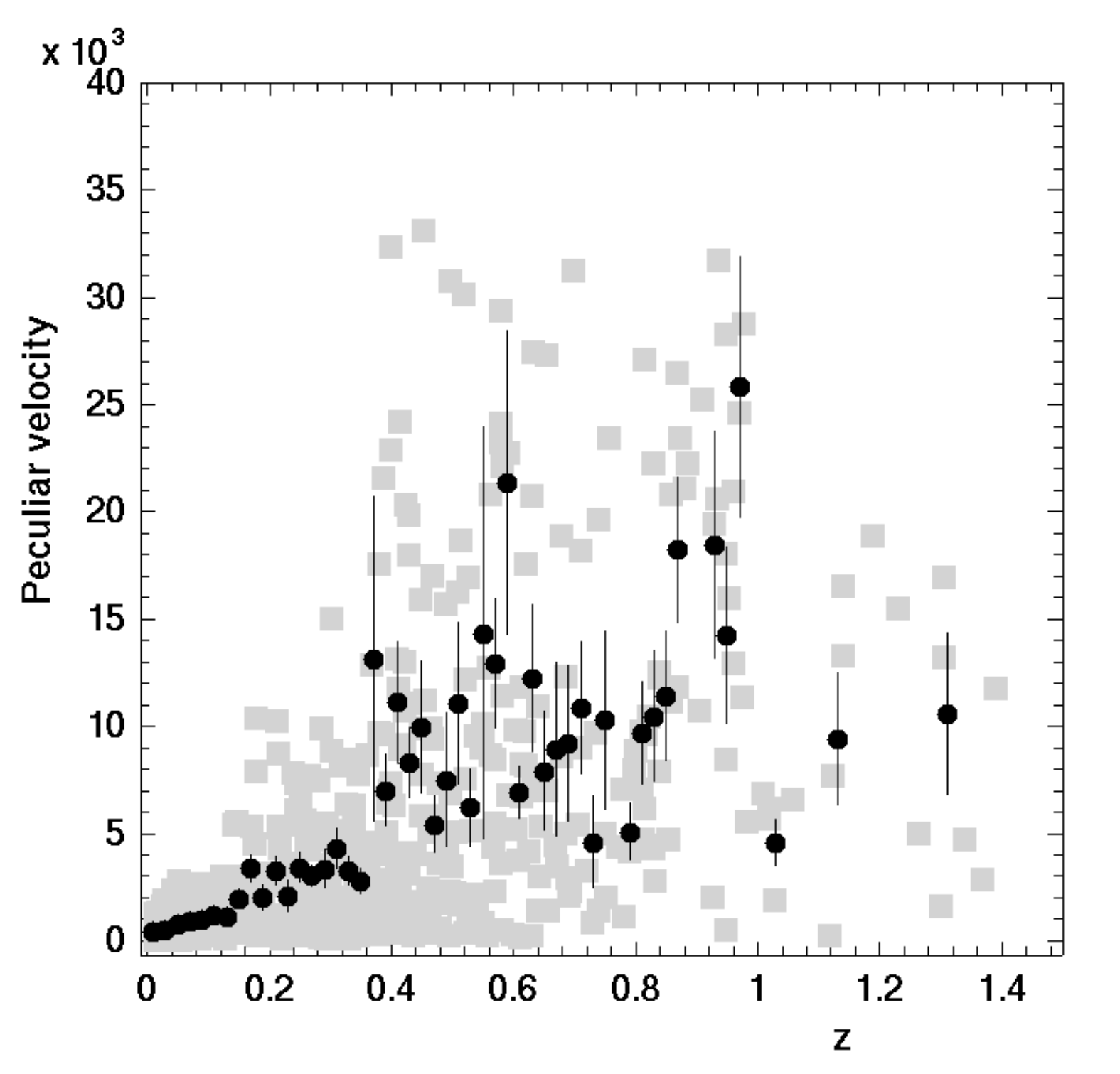,width=7.7cm}}
\caption{{Similar to Fig. \ref{2plot} but the velocities (km/s) are computed from the uncertainties on
luminosity measurements rather than the luminosity residuals $\mu-\bar{\mu}$.}}
\label{vpunc}
\end{figure}

Figure \ref{2plot} shows the absolute value of the computed (radial) peculiar velocity versus redshift for SNe Ia
from Union 2 catalog. We have ignored any probable lensing effect.
The velocities are hence computed from relations (\ref{muzkv}), (\ref{kappav}) and (\ref{zp}).
As can be illustrated, the velocities are $\sim$ 1000 km/s for $z<0.15$
and then increase dramatically with $z$ for higher redshifts.
Figure \ref{vpunc} shows the peculiar velocities produced by the uncertainties
of the observed distance moduli for Union 2 catalog. This is done
by replacing $\Delta\mu(z)$ in relation (\ref{muzkv}) by the uncertainties on the SNe distance moduli
and computing the corresponding  velocity for each SN.
In both figures, each grey box represents a SN Ia. Black circles show the averaged velocities in redshift bin of 0.01.
The error bars correspond to one sigma scatter per bin.  We should mention that the
number of SNe Ia per bin decreases by redshift.
As can be seen, both figures \ref{2plot} and \ref{vpunc} show the same trends.
This is equivalent to say that the uncertainties on SNe Ia luminosity are still large-enough
that the deviations from the background ($\Lambda$CDM) model are not statistically significant.
This fact prevents us from attributing the velocity excess at higher redshifts to the lensing effect.
Furthermore, even for $z < $0.15, the velocities produced by
pure uncertainties on the observed distance moduli are larger than the one produced by $\Delta\mu(z)$.
This shows that the peculiar velocity measurements are not sufficiently precise due to,
once again, the large uncertainties on SNe Ia luminosity measurement.

\section{Conclusion and future prospects}

\label{Sec3}
{{Considering the FRW metric and the $\Lambda$CDM model, one can obtain the luminosity distance as a function
of the cosmological redshift. At low redshifts, the deviation of the observed luminosity distance from the model prediction
is attributed to the source peculiar velocity induced by local structures around the source.  The peculiar velocity is obtained
by computing the amount of difference in the luminosity distance in the $\Lambda$CDM model.
The peculiar velocity measurements are in the heart of modern cosmology to probe the local universe and also to study the cosmological models in large scales.
In this work we investigate the peculiar velocity measurement in perturbed-FRW universe and we assert and show that relativistic corrections are important in our results and interpretation. In one hand, one should calculate the velocities with a relativistic doppler effect, on the other hand we should take into account the effects of lensing, Sachs-Wolfe (the main and integrated one) and the peculiar velocity of the source and observer. In this work we investigate the interplay of the peculiar velocity and lensing at low and specially at intermediate redshifts. We also want to bring the attention of the community to the fact that using less model-dependent methods, such as linear theory, can help us to derive the peculiar velocity by considering the matter distribution around the source. Such methods lead us to reexamine the cosmological models according to
model predictions for the peculiar velocities \cite{Baghram:2014qja}.
In this work, we showed that as we approach to intermediate redshifts, the lensing effect of the line-of-sight structures gets stronger. The deviations of
the luminosity distance from the model, caused by gravitational convergence, become comparable to the
effect of peculiar velocities. Therefore, to compute the peculiar velocity, one should firstly correct the luminosity distance
for the (de)lensing effect.
We showed that with the present data of the standard candles, using the Union 2 data sample, it is not yet possible to extract the contribution of the
peculiar velocity and the gravitational lensing on the luminosity deviation from the background $\Lambda$CDM model.
This is essentially due to the rather large (systematic) uncertainties in SNe Ia data. These uncertainties prevent us
from precise measurement of the peculiar velocity even at low redshifts.
However, including the effect of cosmic convergence on peculiar
velocities is crucial for future surveys (such as LSST \cite{Abell:2009aa} and  Euclid \cite{Amendola:2012ys}).
These surveys will map cosmic structures up to  intermediate redshifts where gravitational lensing plays an important role. Finally worth to mention again that the extension of this work with tilted coordinates \cite{Tsagas:2015mua} can have new prospects on the relativistic effects study in large scale structure observations. }}

\section{Acknowledgements}

We would like to thank Joseph Silk, Roya Mohayaee and Marc Moniez
for their insightful comments and useful discussions.
We thank R. Mansouri and S.M. Mahdavi for facilitating the stay of FH at Sharif University of Technology.
The research of FH  at IAP is supported by a grant from the International Balzan Foundation
via the Oxford New College-Johns Hopkins University Center for Cosmological Studies.

\end{document}